%% file: qcsrm_main.tex
\documentclass[%
 reprint,
 amsmath,amssymb,
 aps,
]{revtex4-2}

\usepackage{graphicx}
\usepackage{dcolumn}
\usepackage{bm}

\begin{document}

\preprint{APS/123-QED}

\title{Quantum computation:\\ Efficient network partitioning for large scale critical infrastructures}

\author{Saikat Ray Majumder}
 \altaffiliation{GE Research, 1 Research Circle, Niskayuna, NY 12309, USA}
\author {Annarita Giani}
 \altaffiliation{GE Research, 1 Research Circle, Niskayuna, NY 12309, USA}
\author {Weiwei Shen}
 \altaffiliation{GE Research, 1 Research Circle, Niskayuna, NY 12309, USA}
\author {Bogdan Neculaes}
 \altaffiliation{GE Research, 1 Research Circle, Niskayuna, NY 12309, USA}
\author{Daiwei Zhu}
\altaffiliation{IonQ Inc, 4505 Campus Dr, College Park, MD 20740, USA}
\author{Sonika Johri}
\altaffiliation{IonQ Inc, 4505 Campus Dr, College Park, MD 20740, USA}

\date{\today}

\begin{abstract}
Quantum computers are emerging as a viable alternative to tackle certain computational problems that are challenging for classical computers. With the rapid development of quantum hardware such as those based on trapped ions, there is practical motivation for identifying risk management problems that are efficiently solvable with these systems. Here we focus on network partitioning as a means for analyzing risk in critical infrastructures and present a quantum approach for its implementation. It is based on the potential speedup quantum computers can provide in the identification of eigenvalues and eigenvectors of sparse graph Laplacians, a procedure which is constrained by time and memory on classical computers.

\end{abstract}

\maketitle


\input{Introduction.tex}
\input{Scrm.tex}
\input{Calg.tex}
\input{Qalg.tex}

\input{poc.tex}
\input{conc.tex}

\bibliography{qcsrm_main}

\end{document}

%% file: Introduction.tex
\section{\label{sec:intro}{Introduction}}

Complex networks are ubiquitous. Systems like power grids, the World Wide Web, social interactions, locomotive and airline networks, cellular networks, food webs, and sensor networks can all be modeled as complex networks. Additionally, in this current era of Industrial Internet of Things, more and more assets are continuously getting connected to each other resulting in large, complex and dynamic networks. The heightened connectivity leads to increased efficiency but often comes at the cost of increased vulnerability. Therefore, it is, important to closely monitor these networks, anticipate and prepare for disruptions and quickly identify efficient mitigation strategies. However, given the size and dynamic nature of these networks, traditional approaches based on discrete optimizations and statistical predictions often face significant limitations. To circumvent some of the limitations in the current modeling techniques, it turns out one can leverage the community structure of the networks. In addition, these network communities also provide a low dimensional graph embedding which can be used in many machine learning applications. 

A traditional method used for community detection is network partitioning. There are existing classical algorithms for network partitioning but the computational and time complexity of such algorithms can grow significantly for large graphs. In this work, we briefly discuss how the rapidly developing technology of quantum computing may provide an edge over classical methods.

%% file: Scrm.tex
\section{\label{sec:real_network}{Networks in the real world}}
Networks in the real world, such as power grids, supply delivery networks and social networks exhibit a high level of order and organization. The degree distribution in such networks often follows a power law in the tail, denoting the fact that many vertices with low degrees coexist with a few vertices with large degrees. These networks exhibit many interesting structural properties, especially when they are large scale and grow in a decentralized and independent fashion, thus not the result of a global, but rather of many local autonomous designs. We briefly describe here two examples of such critical infrastructures, where network analysis can provide significant benefits: supply chain and power grid.

\subsection{\label{sec:scr}{Supply Chain Risk and Resilience}}

\begin{figure}[b]
\includegraphics[width=0.75\columnwidth]{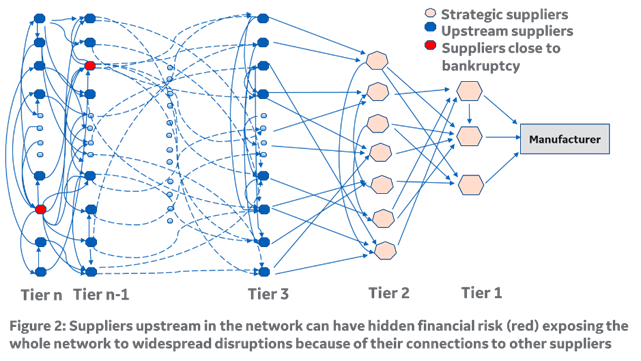}
\caption{\label{fig:scn} A multi-echelon supply chain}
\end{figure}

Suppliers in a supply chain are divided according to the distance to the final product. Tier 1 suppliers provide product to the manufacturer directly. Tier 3 suppliers are two steps down in the chain. Traditional supply chain risk management focuses on Tier 1 suppliers of “critical” goods; however, risk can lie in any tier or echelon of a supply chain \cite{yan2015}. Suppose a lesser-known supplier, several tiers deep in the supply chain, goes out of business due to a lack of working capital availability (red nodes in Fig.~\ref{fig:scn}). This bankruptcy then leads to a cascading disruption in the supply chain due to this company’s structural position in the extended network, ultimately disrupting or shutting down the manufacturing facility of a major OEM (Original Equipment Manufacturer). 

 One such example is Evonik Industries, a little-known raw materials supplier, whose plant explosion in 2012 caused major disruptions in the production of automobiles throughout the global automotive industry~\cite{evonik}. Identifying and mitigating these types of disruption risks is difficult since many such critical suppliers can be several tiers deep in the supply chain and hence not visible to risk managers until the disruption is already occurring. In recent years, techniques from the domains of graph theory and complex network analysis (CN) have been adapted to address such problems and quantify systemic risks and resilience of the supply network in a scalable fashion. This approach may enable risk managers to understand the indirect effects that interventions in one part of the supply chain can have on another part \cite{ritter2004}.

Graph analytics exploit network topology to define properties such as centrality measures, clusters, critical nodes, tipping points and resilience. Risk managers can use these features to gain insights into the nature of the network and be proactive in taking early mitigation steps to address risks at their nascent stages. For instance, this framework can rank suppliers who are more central to the network and should be monitored more closely. These well-connected suppliers play a major role in the network by controlling the overall performance of the network and ensuring a system wide coordination to drive greater efficiency. Due to their high connectivity, these hub firms have an outsized influence over the network, which leads to better self-coordination, less duplications and lower transaction costs. One can measure the impact that a supplier has on the efficiency of a network by calculating the supplier’s contribution to the characteristic path of the network. A network with short characteristic path length will ensure quick diffusion of new information enabling more efficient material and financial flows throughout the network. If suppliers default and are removed from a network, the characteristic path lengths will increase, and ultimately vertex pairs will become disconnected and communication between them through the network will become impossible. One can develop metrics of rapid change to signal that the supply network is approaching a tipping point. In many networks tipping points exist at which dynamics of the network abruptly changes. War, riots, pandemic, natural disaster, or economic downturn are obvious triggers of such tipping points. Yet, not all networks succumb to such exogenous shocks. One can investigate how stronger financial health of the suppliers can make the network more resilient to external risks.

\subsection{\label{sec:grid}{Power Grid}}
Power grid is a highly complex cyber-physical system with lots of interconnected components. Physical measurement data are delivered from remote technical units (RTUs) to supervisory control and data acquisition (SCADA) systems and then to Advanced Energy Management System (AEMS) applications responsible for controlling and monitoring the power system. This gives rise to a significant challenge in maintaining and operating the grid while ensuring high level of resiliency against normal disruptions and cyber attacks.

Graph theory provides a mathematical object that naturally encodes relationships and hence provides a robust framework to build such applications ~\cite{PhysRevResearch.3.023161,doi:10.1063/1.5092629,Szoplik2010a, Szoplik2010b}. For instance, with the data cast as a graph, the problem often boils down to identifying a small subset of nodes with much higher volume of network traffic, than is typical for those nodes, indicating the onset of some malicious activity.  Essentially the goal it to identify network interactions which do not fit the model of typical, normal behavior and thereby detect and counter malicious activity. But identifying graph patterns from within the vast and complex network is a classic subgraph isomorphism problem and is known to be computationally expensive and NP-complete ~\cite{Cook_1971, Cormen_1990}.
Additional complexity is the requirement to detect the pattern before it is fully instantiated. This introduces new algorithmic challenges because one cannot afford to index a dynamic graph frequently enough for applications with real-time constraints.

%% file: Calg.tex
\section{\label{sec:calg}{Classical approach}}
In both of the above use cases (and, also in similar other application domains) the primary challenge is the scalability of the traditional methodologies. These networks are dynamic complex systems with non-linear interactions and often need to be analyzed at a system level. However, the networks can comprise of tens of thousand of nodes and that is where many traditional methods run into computational challenges. One potential solution is to find appropriate clusters, or communities in these networks and thereby reduce the dimensionality of the problem by partitioning the large graph into smaller sub-graphs.

\subsection{Community Detection}
 Large networks exhibit lack of homogeneity both globally and locally. The local inhomogeneities give rise to a dense concentration of edges within groups of vertices and very sparse connections between groups. This feature of a network is called its community structure or clustering. Communities reveal the hierarchical organization of the network and mark groups of nodes which share common properties, exchange more transactions or information or have similar functions \cite{lu2018, santo2010}. Community detection is therefore a very important task in network analysis.

The presence of communities in real world networks is quite intuitive. However, the task of detecting these communities is often very challenging. One problem is that the definitions of a community and a partition are not rigorous. Classical techniques for data clustering, like hierarchical, partitional and spectral clustering have been adopted for graph clustering. Other methods include neural network clustering and multi dimensional scaling techniques, such as singular value decomposition and principal component analysis. Many of these clustering techniques are NP-hard.  

Spectral clustering uses the graph Lapalacian. Normal graph Lapalacian is defined as follows:
\begin{equation}
    L(G) = D(G) - A(G)
\end{equation}
where, $A$ is the adjacency matrix of the graph $G$ and $D$ is the degree matrix. The Laplacian is positive semidefinite, that is, all eigenvalues are non-negative. Eigenvector decomposition of the Laplacian is closely related to the clustering problem. The number of zero eigenvalues correspond to the number of connected components in the graph. Eigenvalues close to zero denote that there is almost a separation into two components. Hence, if there are $N_c$ clusters in a network, in spectral clustering it is required to find the eigenvectors of the Laplacian corresponding to the smallest $N_c$ eigenvalues. The second smallest eigenvalue of the Lapalacian is called the Fiedler eigenvalue and the corresponding eigenvector is called the Fiedler vector. Fiedler value indicates how well connected the graph is and Fiedler vector can be used to bisect the graph based on the sign of the corresponding element in the vector.    

For large graphs with $N$ vertices it is however impossible to have exact diagonalization solutions as the time complexity is $O(N^3)$. In such cases approximate algorithms are used \cite{santo2010}. As outlined in \cite{santo2010}, approximate algorithms, including those for sparse graphs cannot scale faster than $O(N)$ or $O(M)$ (where $M$ is the number of edges in the graph). Of even more serious concern may be the memory requirements for diagonalization which also scale as $O(N)$. Hence, for large graphs, even approximate algorithms on classical computers may be insufficient to diagonalize the graph Laplacian as they will scale at least linearly in the number of vertices and edges of the graph. In these cases, the rapidly developing technology of quantum computing may provide an edge over classical methods.

%% file: Qalg.tex
\section{\label{sec:qalg}{Quantum approach}}
Quantum computers work with quantum bits, or `qubits', which differ from classical bits in that they can be in a superposition of 0 and 1 at the same time. Further, qubits can be entangled, so that a system of $n$ qubits can be in a superposition of $2^n$ classical states (bit strings) $k$ described by the quantum state $|\phi\rangle=\sum_k a_k |k\rangle$. Here $a_k$ are complex numbers which satisfy the rule $\sum_k |a_k|^2=1$.

On quantum computers, the problem of finding eigenvalues of a Hermitian matrix can be tackled using the quantum phase estimation algorithm, which proceeds as follows: given a unitary matrix $U$ and a quantum state $|\psi\rangle$ defined on $n=\log_2(N)$ qubits such that $U|\psi\rangle=e^{i2\pi\theta}|\psi\rangle$, phase estimation allows one to determine $\theta$ with precision $\delta$ using $O(\log(1/\delta)+\log(N))$ qubits with $O(1/\delta)$ controlled applications of the unitary matrix $U$. $U$ can be expressed as the `time-evolution' under the Hermitian matrix. In the problem of (undirected) graph partitioning, the Laplacian $L$ being real and symmetric is Hermitian, so we can write $U=e^{iLt}$. Let's assume that $L$ is normalized such that its maximum eigenvalue is 1 and we set $t=2\pi$. Then $\delta$ is chosen such that it is the smaller of the distance between the eigenvalues of interest and the precision to which an eigenvalue needs to be known.  For quantum phase estimation to be effectively applied, one must then find an efficient implementation of $U$, as well as an efficient way to prepare the quantum state $\psi$. 

We first outline the task of implementing $U=e^{iLt}$. A technique for this is given in Ref. \cite{low2019}. According to this method, given a $d$-sparse Hermitian matrix $L$ (which is normalized such that $\frac{||L||}{d||L||_{max}}=1$), one can implement the operator $e^{iLt}$ (up to an error $\epsilon$) with $O(t+\log(1/\epsilon))$ calls to an oracle that returns the matrix element given the row and column, and an oracle that returns a sequence of column indices in a particular row. These oracles will typically be implemented in time $O(\log(N))$. If $L$ is sparse, as is typical for practical cases of interest, the application of $L$ with sparsity $d$ will have time complexity to leading order of $O(d\log(N))$, giving an overall runtime that scales as $O(d\log(N)/\delta)$. Thus, the quantum algorithm provides exponential speed-up in the size of the matrix for both time and memory.




More speculatively, the time complexity may be further reduced if as a preprocessing step, a variational quantum algorithm is used to learn a quantum circuit which encodes time-evolution under the graph Laplacian. While this is a heuristic procedure for which time-complexity scaling is not guaranteed, it can presumably lead to finding a more efficient implementation of the time-evolution operator. This should form an area of research for risk management on near-term quantum computers.

Next, we turn to the task of preparing the initial state $|\psi_0\rangle$ on which $U$ will act. Since we don't know the eigenvectors in advance, it is not possible to prepare an exact version of $|\psi\rangle$ even if we knew how to do it efficiently. Therefore, our goal is to prepare $|\psi_0\rangle$ as close to $|\psi\rangle$ as possible. On repeating the phase estimation procedure several times, a distribution over the eigenvalues will be obtained, where the probability of obtaining a particular value is equal to $|\langle\psi_j|\psi_0\rangle|^2$. In principle, starting with a random input state will give some non-zero overlap with the desired largest eigenvectors, but these may be too small to be practically useful. Hence a few different strategies can be adopted:

1. Using matrix product states which scale as $\log(n)$ to prepare a bounded-entanglement approximation of the largest eigenvalue state, and converting this to a quantum circuit.

2. Adiabatic approach: This involves starting in a quantum state that is a product state of the qubits, or one that can be prepared with a low-depth circuit. This starting state is the ground state of a known Hamiltonian whose time-evolution is easily implementable. Then a discretized adiabatic evolution which slowly changes the time-evolution from the starting Hamiltonian to that under $L$ can be used to prepare an approximation of the ground states of $L$. The time for this approach scales as $1/\delta$.

3. Variational approach: A heuristic approach which uses a variational quantum circuit whose parameters are tuned according to a cost function based on $L$

In addition to the eigenvalues, eigenvectors can also be determined by sampling the output eigenstate. The probability
of measuring a particular basis state $k$ is $|a_k|^2$, where $a_k$ is the eigenvector element. Therefore, the largest elements of the eigenvector can be determined efficiently. More precisely, the eigenvector elements can be determined to a precision $1/\sqrt{N_{\text{samples}}}$, where $N_{\text{samples}}$ is the number of times the procedure is repeated. 

The number of 0 eigenvalues can be determined by preparing multiple copies of $|\psi_0\rangle$ starting from orthogonal initializations of the qubits and then projecting them into a state which gives 0 eigenvalue after phase estimation. The number of unique states that can be so prepared then gives the degeneracy of the eigenvalue.

%% file: poc.tex
\section{\label{sec:poc}{Realization on Quantum Hardware}}
Quantum hardware, while still in a nascent stage, is rapidly advancing to be powerful enough to demonstrate algorithms like the ones described above. Leading quantum hardware platforms include trapped ions, superconducting qubits, neutral atoms, and photonic qubits. With the rapid development of quantum computing hardware, proposals for benchmarking performance in an application-oriented manner have been put forth \cite{qedc}. One such example is Algorithmic Qubits (AQ) \footnote{{https}://ionq.com/posts/february-23-2022-algorithmic-qubits}. Under this definition, quantum hardware from IonQ has advanced from AQ 6 to AQ 23 in 2 years, and is projected to reach AQ 64 by 2025, at which point it will be beyond the simulation capabilities of classical computers.

Small scale demonstrations of quantum phase estimation algorithms discussed in this white paper include one on a silicon photonic chip \cite{photonic} and using machine learning to enhance the measurement of eigenvalues \cite{ml}.

%% file: conc.tex
\section{\label{sec:conc}{Conclusions and Proposed Future Work}}
Large scale complex networks are key for today’s world, with multiple national security implications. However, the large sizes of these networks often limit the use of standard algorithms and approaches to analyze them. Community structure of the networks provides a powerful feature to circumvent some of these challenges. Since there is an expected low level of interdependence between the communities, the ensuing analysis more naturally renders to parallel computation thereby making it scalable and more efficient. For instance, identifying clusters of suppliers based on industrial sectors or regions enables a supply chain risk manager to better understand the risk dynamics and their inter-dependencies while simultaneously reducing the computational burden of analysing the full supply delivery network.

Network partitioning is a popular technique in community detection and can be done by diagonalizing the graph Laplacian. However, this approach is constrained by time and memory on classical computers. Quantum computing can identify eigenvalues and eigenvectors for sparse matrices exponentially faster in the size of the matrix compared to classical computers and thus has applications in risk management of networks. 

While quantum computing is a nascent technology, the quality and robustness of quantum computers is improving rapidly. We propose the following research strategy for pursuing this approach:\\ 
- Identify examples of networks that are relevant for critical infrastructure\\
- Develop concrete quantum algorithms customized for these networks and implement them in a quantum software framework\\
- Carry out resource estimates for the number of qubits and fidelity required to analyze real-life networks\\
- Test the algorithms on simulators to verify correctness and robustness to noise\\
- Test simplified versions of these algorithms on available quantum hardware

At the conclusion of the vision detailed above, one would be able to quantify the impact of quantum computing on network partitioning, a computing problem with dramatic civilian and national security implications. In addition, the effort will lay out the hardware timeline for practical implementation of quantum solutions to this problem.

%% file: qcsrm_main.bbl
\providecommand{\noopsort}[1]{}\providecommand{\singleletter}[1]{#1}%
\begin{thebibliography}{16}%
\makeatletter
\providecommand \@ifxundefined [1]{%
 \@ifx{#1\undefined}
}%
\providecommand \@ifnum [1]{%
 \ifnum #1\expandafter \@firstoftwo
 \else \expandafter \@secondoftwo
 \fi
}%
\providecommand \@ifx [1]{%
 \ifx #1\expandafter \@firstoftwo
 \else \expandafter \@secondoftwo
 \fi
}%
\providecommand \natexlab [1]{#1}%
\providecommand \enquote  [1]{``#1''}%
\providecommand \bibnamefont  [1]{#1}%
\providecommand \bibfnamefont [1]{#1}%
\providecommand \citenamefont [1]{#1}%
\providecommand \href@noop [0]{\@secondoftwo}%
\providecommand \href [0]{\begingroup \@sanitize@url \@href}%
\providecommand \@href[1]{\@@startlink{#1}\@@href}%
\providecommand \@@href[1]{\endgroup#1\@@endlink}%
\providecommand \@sanitize@url [0]{\catcode `\\12\catcode `\$12\catcode
  `\&12\catcode `\#12\catcode `\^12\catcode `\_12\catcode `\%12\relax}%
\providecommand \@@startlink[1]{}%
\providecommand \@@endlink[0]{}%
\providecommand \url  [0]{\begingroup\@sanitize@url \@url }%
\providecommand \@url [1]{\endgroup\@href {#1}{\urlprefix }}%
\providecommand \urlprefix  [0]{URL }%
\providecommand \Eprint [0]{\href }%
\providecommand \doibase [0]{https://doi.org/}%
\providecommand \selectlanguage [0]{\@gobble}%
\providecommand \bibinfo  [0]{\@secondoftwo}%
\providecommand \bibfield  [0]{\@secondoftwo}%
\providecommand \translation [1]{[#1]}%
\providecommand \BibitemOpen [0]{}%
\providecommand \bibitemStop [0]{}%
\providecommand \bibitemNoStop [0]{.\EOS\space}%
\providecommand \EOS [0]{\spacefactor3000\relax}%
\providecommand \BibitemShut  [1]{\csname bibitem#1\endcsname}%
\let\auto@bib@innerbib\@empty
\bibitem [{\citenamefont {Yan}\ \emph {et~al.}(2015)\citenamefont {Yan},
  \citenamefont {Choi}, \citenamefont {Kim},\ and\ \citenamefont
  {Yang}}]{yan2015}%
  \BibitemOpen
  \bibfield  {author} {\bibinfo {author} {\bibfnamefont {T.}~\bibnamefont
  {Yan}}, \bibinfo {author} {\bibfnamefont {T.}~\bibnamefont {Choi}}, \bibinfo
  {author} {\bibfnamefont {Y.}~\bibnamefont {Kim}},\ and\ \bibinfo {author}
  {\bibfnamefont {Y.}~\bibnamefont {Yang}},\ }\bibfield  {title} {\bibinfo
  {title} {A theory of the nexus supplier: A critical supplier from a network
  perspective},\ }\href@noop {} {\bibfield  {journal} {\bibinfo  {journal} {J
  Supply Chain Manag}\ }\textbf {\bibinfo {volume} {51}},\ \bibinfo {pages}
  {52} (\bibinfo {year} {2015})}\BibitemShut {NoStop}%
\bibitem [{evo(2012)}]{evonik}%
  \BibitemOpen
  \bibfield  {title} {\bibinfo {title} {Explosion at evonik factory may have
  serious knock-on effect for global car production},\ }\href@noop {}
  {\bibfield  {journal} {\bibinfo  {journal} {Automotive Industries AI.}\
  }\textbf {\bibinfo {volume} {192}},\ \bibinfo {pages} {4} (\bibinfo {year}
  {2012})}\BibitemShut {NoStop}%
\bibitem [{\citenamefont {Ritter}\ \emph {et~al.}(2004)\citenamefont {Ritter},
  \citenamefont {Wilkinson},\ and\ \citenamefont {Johnston}}]{ritter2004}%
  \BibitemOpen
  \bibfield  {author} {\bibinfo {author} {\bibfnamefont {T.}~\bibnamefont
  {Ritter}}, \bibinfo {author} {\bibfnamefont {I.}~\bibnamefont {Wilkinson}},\
  and\ \bibinfo {author} {\bibfnamefont {W.}~\bibnamefont {Johnston}},\
  }\bibfield  {title} {\bibinfo {title} {Managing in complex business
  networks},\ }\href@noop {} {\bibfield  {journal} {\bibinfo  {journal}
  {Industrial Marketing Management}\ }\textbf {\bibinfo {volume} {33}},\
  \bibinfo {pages} {175} (\bibinfo {year} {2004})}\BibitemShut {NoStop}%
\bibitem [{\citenamefont {Kaiser}\ and\ \citenamefont
  {Witthaut}(2021)}]{PhysRevResearch.3.023161}%
  \BibitemOpen
  \bibfield  {author} {\bibinfo {author} {\bibfnamefont {F.}~\bibnamefont
  {Kaiser}}\ and\ \bibinfo {author} {\bibfnamefont {D.}~\bibnamefont
  {Witthaut}},\ }\bibfield  {title} {\bibinfo {title} {Topological theory of
  resilience and failure spreading in flow networks},\ }\href
  {https://doi.org/10.1103/PhysRevResearch.3.023161} {\bibfield  {journal}
  {\bibinfo  {journal} {Phys. Rev. Research}\ }\textbf {\bibinfo {volume}
  {3}},\ \bibinfo {pages} {023161} (\bibinfo {year} {2021})}\BibitemShut
  {NoStop}%
\bibitem [{\citenamefont {Guo}\ \emph {et~al.}(2019)\citenamefont {Guo},
  \citenamefont {Yu}, \citenamefont {Iu}, \citenamefont {Fernando},\ and\
  \citenamefont {Zheng}}]{doi:10.1063/1.5092629}%
  \BibitemOpen
  \bibfield  {author} {\bibinfo {author} {\bibfnamefont {H.}~\bibnamefont
  {Guo}}, \bibinfo {author} {\bibfnamefont {S.~S.}\ \bibnamefont {Yu}},
  \bibinfo {author} {\bibfnamefont {H.~H.~C.}\ \bibnamefont {Iu}}, \bibinfo
  {author} {\bibfnamefont {T.}~\bibnamefont {Fernando}},\ and\ \bibinfo
  {author} {\bibfnamefont {C.}~\bibnamefont {Zheng}},\ }\bibfield  {title}
  {\bibinfo {title} {A complex network theory analytical approach to power
  system cascading failure—from a cyber-physical perspective},\ }\href
  {https://doi.org/10.1063/1.5092629} {\bibfield  {journal} {\bibinfo
  {journal} {Chaos: An Interdisciplinary Journal of Nonlinear Science}\
  }\textbf {\bibinfo {volume} {29}},\ \bibinfo {pages} {053111} (\bibinfo
  {year} {2019})},\ \Eprint
  {https://arxiv.org/abs/https://doi.org/10.1063/1.5092629}
  {https://doi.org/10.1063/1.5092629} \BibitemShut {NoStop}%
\bibitem [{\citenamefont {Szoplik}(2010{\natexlab{a}})}]{Szoplik2010a}%
  \BibitemOpen
  \bibfield  {author} {\bibinfo {author} {\bibfnamefont {J.}~\bibnamefont
  {Szoplik}},\ }\bibfield  {title} {\bibinfo {title} {Quantitative analysis of
  the heterogeneity for gas flow in the pipeline system.},\ }\href@noop {}
  {\bibfield  {journal} {\bibinfo  {journal} {Gaz, Woda i Technika Sanitarna}\
  }\textbf {\bibinfo {volume} {1}} (\bibinfo {year}
  {2010}{\natexlab{a}})}\BibitemShut {NoStop}%
\bibitem [{\citenamefont {Szoplik}(2010{\natexlab{b}})}]{Szoplik2010b}%
  \BibitemOpen
  \bibfield  {author} {\bibinfo {author} {\bibfnamefont {J.}~\bibnamefont
  {Szoplik}},\ }\bibfield  {title} {\bibinfo {title} {The application of the
  graph theory to the analysis of gas flow in a pipeline network.},\
  }\href@noop {} {\bibfield  {journal} {\bibinfo  {journal} {37th International
  Conference of SSCHE}\ }\textbf {\bibinfo {volume} {CD-ROM}} (\bibinfo {year}
  {2010}{\natexlab{b}})}\BibitemShut {NoStop}%
\bibitem [{\citenamefont {Cook}(1971)}]{Cook_1971}%
  \BibitemOpen
  \bibfield  {author} {\bibinfo {author} {\bibfnamefont {S.~A.}\ \bibnamefont
  {Cook}},\ }\bibfield  {title} {\bibinfo {title} {The complexity of
  theorem-proving procedures},\ }\href@noop {} {\bibfield  {journal} {\bibinfo
  {journal} {Proc. 3rd ACM Symposium on Theory of Computing}\ } (\bibinfo
  {year} {1971})}\BibitemShut {NoStop}%
\bibitem [{\citenamefont {T.~H.~Cormen}\ and\ \citenamefont
  {Rivest}(1990)}]{Cormen_1990}%
  \BibitemOpen
  \bibfield  {author} {\bibinfo {author} {\bibfnamefont {C.~E.~L.}\
  \bibnamefont {T.~H.~Cormen}}\ and\ \bibinfo {author} {\bibfnamefont {R.~L.}\
  \bibnamefont {Rivest}},\ }\href@noop {} {\emph {\bibinfo {title}
  {Introduction to Algorithms}}}\ (\bibinfo  {publisher} {Cambridge, Mass.: MIT
  Press},\ \bibinfo {year} {1990})\BibitemShut {NoStop}%
\bibitem [{\citenamefont {Lu}\ \emph {et~al.}(2018)\citenamefont {Lu},
  \citenamefont {Wahlström},\ and\ \citenamefont {Nehorai}}]{lu2018}%
  \BibitemOpen
  \bibfield  {author} {\bibinfo {author} {\bibfnamefont {Z.}~\bibnamefont
  {Lu}}, \bibinfo {author} {\bibfnamefont {J.}~\bibnamefont {Wahlström}},\
  and\ \bibinfo {author} {\bibfnamefont {A.}~\bibnamefont {Nehorai}},\
  }\bibfield  {title} {\bibinfo {title} {Community detection in complex
  networks via clique conductance},\ }\href@noop {} {\bibfield  {journal}
  {\bibinfo  {journal} {Scientific Reports}\ }\textbf {\bibinfo {volume} {8}}
  (\bibinfo {year} {2018})}\BibitemShut {NoStop}%
\bibitem [{\citenamefont {Fortunato}(2010)}]{santo2010}%
  \BibitemOpen
  \bibfield  {author} {\bibinfo {author} {\bibfnamefont {S.}~\bibnamefont
  {Fortunato}},\ }\bibfield  {title} {\bibinfo {title} {Community detection in
  graphs},\ }\href@noop {} {\bibfield  {journal} {\bibinfo  {journal} {Physics
  Reports}\ }\textbf {\bibinfo {volume} {486}},\ \bibinfo {pages} {75}
  (\bibinfo {year} {2010})}\BibitemShut {NoStop}%
\bibitem [{\citenamefont {Low}\ and\ \citenamefont {Chuang}(2019)}]{low2019}%
  \BibitemOpen
  \bibfield  {author} {\bibinfo {author} {\bibfnamefont {G.~H.}\ \bibnamefont
  {Low}}\ and\ \bibinfo {author} {\bibfnamefont {I.~L.}\ \bibnamefont
  {Chuang}},\ }\bibfield  {title} {\bibinfo {title} {Hamiltonian simulation by
  qubitization},\ }\href@noop {} {\bibfield  {journal} {\bibinfo  {journal}
  {Quantum}\ }\textbf {\bibinfo {volume} {3}} (\bibinfo {year}
  {2019})}\BibitemShut {NoStop}%
\bibitem [{\citenamefont {Lubinski}\ \emph {et~al.}(2021)\citenamefont
  {Lubinski}, \citenamefont {Johri}, \citenamefont {Varosy}, \citenamefont
  {Coleman}, \citenamefont {Zhao}, \citenamefont {Necaise}, \citenamefont
  {Baldwin}, \citenamefont {Mayer},\ and\ \citenamefont {Proctor}}]{qedc}%
  \BibitemOpen
  \bibfield  {author} {\bibinfo {author} {\bibfnamefont {T.}~\bibnamefont
  {Lubinski}}, \bibinfo {author} {\bibfnamefont {S.}~\bibnamefont {Johri}},
  \bibinfo {author} {\bibfnamefont {P.}~\bibnamefont {Varosy}}, \bibinfo
  {author} {\bibfnamefont {J.}~\bibnamefont {Coleman}}, \bibinfo {author}
  {\bibfnamefont {L.}~\bibnamefont {Zhao}}, \bibinfo {author} {\bibfnamefont
  {J.}~\bibnamefont {Necaise}}, \bibinfo {author} {\bibfnamefont {C.~H.}\
  \bibnamefont {Baldwin}}, \bibinfo {author} {\bibfnamefont {K.}~\bibnamefont
  {Mayer}},\ and\ \bibinfo {author} {\bibfnamefont {T.}~\bibnamefont
  {Proctor}},\ }\href {https://doi.org/10.48550/ARXIV.2110.03137} {\bibinfo
  {title} {Application-oriented performance benchmarks for quantum computing}}
  (\bibinfo {year} {2021})\BibitemShut {NoStop}%
\bibitem [{Note1()}]{Note1}%
  \BibitemOpen
  \bibinfo {note}
  {{https}://ionq.com/posts/february-23-2022-algorithmic-qubits}\BibitemShut
  {NoStop}%
\bibitem [{\citenamefont {Paesani}\ \emph {et~al.}(2017)\citenamefont
  {Paesani}, \citenamefont {Gentile}, \citenamefont {Santagati}, \citenamefont
  {Wang}, \citenamefont {Wiebe}, \citenamefont {Tew}, \citenamefont {O'Brien},\
  and\ \citenamefont {Thompson}}]{photonic}%
  \BibitemOpen
  \bibfield  {author} {\bibinfo {author} {\bibfnamefont {S.}~\bibnamefont
  {Paesani}}, \bibinfo {author} {\bibfnamefont {A.~A.}\ \bibnamefont
  {Gentile}}, \bibinfo {author} {\bibfnamefont {R.}~\bibnamefont {Santagati}},
  \bibinfo {author} {\bibfnamefont {J.}~\bibnamefont {Wang}}, \bibinfo {author}
  {\bibfnamefont {N.}~\bibnamefont {Wiebe}}, \bibinfo {author} {\bibfnamefont
  {D.~P.}\ \bibnamefont {Tew}}, \bibinfo {author} {\bibfnamefont {J.~L.}\
  \bibnamefont {O'Brien}},\ and\ \bibinfo {author} {\bibfnamefont {M.~G.}\
  \bibnamefont {Thompson}},\ }\bibfield  {title} {\bibinfo {title}
  {Experimental bayesian quantum phase estimation on a silicon photonic chip},\
  }\href {https://doi.org/10.1103/PhysRevLett.118.100503} {\bibfield  {journal}
  {\bibinfo  {journal} {Phys. Rev. Lett.}\ }\textbf {\bibinfo {volume} {118}},\
  \bibinfo {pages} {100503} (\bibinfo {year} {2017})}\BibitemShut {NoStop}%
\bibitem [{\citenamefont {Lumino}\ \emph {et~al.}(2018)\citenamefont {Lumino},
  \citenamefont {Polino}, \citenamefont {Rab}, \citenamefont {Milani},
  \citenamefont {Spagnolo}, \citenamefont {Wiebe},\ and\ \citenamefont
  {Sciarrino}}]{ml}%
  \BibitemOpen
  \bibfield  {author} {\bibinfo {author} {\bibfnamefont {A.}~\bibnamefont
  {Lumino}}, \bibinfo {author} {\bibfnamefont {E.}~\bibnamefont {Polino}},
  \bibinfo {author} {\bibfnamefont {A.~S.}\ \bibnamefont {Rab}}, \bibinfo
  {author} {\bibfnamefont {G.}~\bibnamefont {Milani}}, \bibinfo {author}
  {\bibfnamefont {N.}~\bibnamefont {Spagnolo}}, \bibinfo {author}
  {\bibfnamefont {N.}~\bibnamefont {Wiebe}},\ and\ \bibinfo {author}
  {\bibfnamefont {F.}~\bibnamefont {Sciarrino}},\ }\bibfield  {title} {\bibinfo
  {title} {Experimental phase estimation enhanced by machine learning},\ }\href
  {https://doi.org/10.1103/PhysRevApplied.10.044033} {\bibfield  {journal}
  {\bibinfo  {journal} {Phys. Rev. Applied}\ }\textbf {\bibinfo {volume}
  {10}},\ \bibinfo {pages} {044033} (\bibinfo {year} {2018})}\BibitemShut
  {NoStop}%
\end{thebibliography}%
